\documentclass[twocolumn,amsmath,amssymb,prb]{revtex4}

\usepackage{graphicx} 
\usepackage{dcolumn}
\usepackage[usenames,dvipsnames]{color}

\usepackage[version=3]{mhchem} 

\definecolor{blue}{rgb}{0.,.0,1.0} 
\def\re#1{{\color{black} {#1}}}

\def \be {\begin{equation}}
\def \ee {\end{equation}}
\def \ben {\begin{eqnarray}}
\def \een {\end{eqnarray}}

\begin{document}
\title{Robust and fragile quantum effects in the transfer kinetics of delocalized excitons between B850 units of LH2 complexes} 

\author{Seogjoo J. Jang}
\email{seogjoo.jang@qc.cuny.edu}
\affiliation{Department of Chemistry and Biochemistry, Queens College, City University of New York, 65-30 Kissena Boulevard, Queens, New York 11367\footnote{mailing address}  \& PhD programs in Chemistry and Physics, and Initiative for the Theoretical Sciences, Graduate Center, City University of New York, 365 Fifth Avenue, New York, NY 10016}
\title{Robust and Fragile Quantum Effects in the Transfer Kinetics of Delocalized Excitons between B850 Units of LH2 Complexes}
\date{Published in the {\it Journal of Physical Chemistry Letters} {\bf 9}, 6576-6583 (2018)}


\begin{abstract}
Aggregates of light harvesting 2 (LH2) complexes form the major exciton-relaying domain  
in the photosynthetic unit of purple bacteria.   Application of a  generalized master equation  to pairs of the B850 units of LH2 complexes, where excitons predominantly reside, provide quantitative information on how the inter-LH2 exciton transfer  depends on the distance, relative rotational angle, and the relative energies of the two LH2s.   The distance dependence demonstrates significant enhancement of the rate due to quantum delocalization of excitons, the qualitative nature of which remains robust against the disorder.   The angle dependence reflects isotropic nature of exciton transfer, which remains similar for the ensemble of disorder.  The variation of the rate on relative excitation energies of LH2 exhibits resonance peaks, which however is fragile as the disorder becomes significant.    Overall, the average transfer times between two LH2s are estimated to be in the range of $4 - 25\ {\rm ps}$ for physically plausible inter-LH2 distances.      
\end{abstract}
\maketitle

While the natural photosynthesis has low energy conversion efficiency overall, light harvesting (LH) processes constituting its initial stage are executed with almost perfect quantum efficiency.\cite{blankenship-2,scholes-jpcl1,scholes-nc3,jang-rmp90}  Decades of spectroscopic and computational studies \cite{blankenship-2,renger-pr343,hu-qrb35,cogdell-qrb39,renger-pr102,jang-wires,cheng-arpc60,jankowiak-cr111,olaya-castro-irpc30,pachon-pccp14,schroter-pr567,chenu-arpc66,konig-cpc13,lee-arpc67,mirkovic-cr117,levi-rpp78,curutchet-cr117,kondo-cr117,jang-rmp90} have shown that Frenkel-type excitons\cite{frenkel-pr37,davydov} are actively utilized in these LH processes.  This is puzzling because Frenkel excitons are fragile and easily disrupted even in the presence of modest amount of disorder and fluctuations.  How natural LH complexes utilize such delicate forms of excitons and create robustly efficient mechanisms of energy transfer is an issue of fundamental importance that has been subject to extensive investigations in recent years.\cite{leegwater-jpc100,jang-rmp90,jang-wires,huo-jpcl2,olbrich-jpcl2,skochdopole-jpcl2,kreisbeck-jpcl3,kassal-jpcl4,chin-np9,ai-jpcl4,kim-jpcl6,lambert-np9,croce-nchembio10,fujita-jpcl3,thyrhaug-jpcl7}  \re{These efforts have contributed significant progress in describing relatively short length scale exciton dynamics confined within a single LH complex.  However,  
how exciton transport occurs across aggregates of multiple LH complexes and how quantum nature of Frenkel-type excitons plays a role in such long range exciton transfer 
still remain poorly understood}.  The photosynthetic unit (PSU) of purple bacteria\cite{hu-qrb35,cogdell-qrb39,sundstrom-jpcb103} is one of the best examples to investigate this issue because of its \re{organizational} simplicity, significant electronic coherence \re{within its constituting LH complexes}, and well-established structural and energetic information.

Under normal light conditions, the PSU of a purple bacterium contains\cite{hu-qrb35,cogdell-qrb39,montemayor-jpcb122} two types of LH antenna pigment-protein complexes referred to as LH1 and LH2.  Of these, LH2 complexes play the primary role of harnessing photons in the form of delocalized excitons.  These excitons are relayed among LH2 complexes towards reaction centers enclosed within LH1 complexes, \re{leading to} 
\re{production of}  charges that are eventually needed for creating proton gradients across the membrane.   Structural and spectroscopic data focusing on individual characteristics of LH2 complexes are relatively well known at present.\cite{hu-qrb35,cogdell-qrb39,montemayor-jpcb122,jang-rmp90}  However, \re{many important issues still need} to be clarified in understanding  how the whole aggregates of LH2 complexes function as a reliable and effective medium for the migration of excitons.  

For example, how does the exciton transfer dynamics between LH2 complexes depend on their relative arrangement such as  inter-LH2 distance and relative orientations?   To what extent does coherent delocalization of excitons contribute to the enhancement of exciton transfer rate?  How can the aggregates of LH2 complexes serve as efficient exciton migration domain without apparent energetic downhill features?  Most importantly, what aspects of quantum effects due to delocalized Frenkel excitons remain robust against the disorder?   Clear answers for these questions will \re{help identifying} the genuine roles of the quantum nature of excitons in the LH processes. 

\begin{figure}
\includegraphics[width=3.5in]{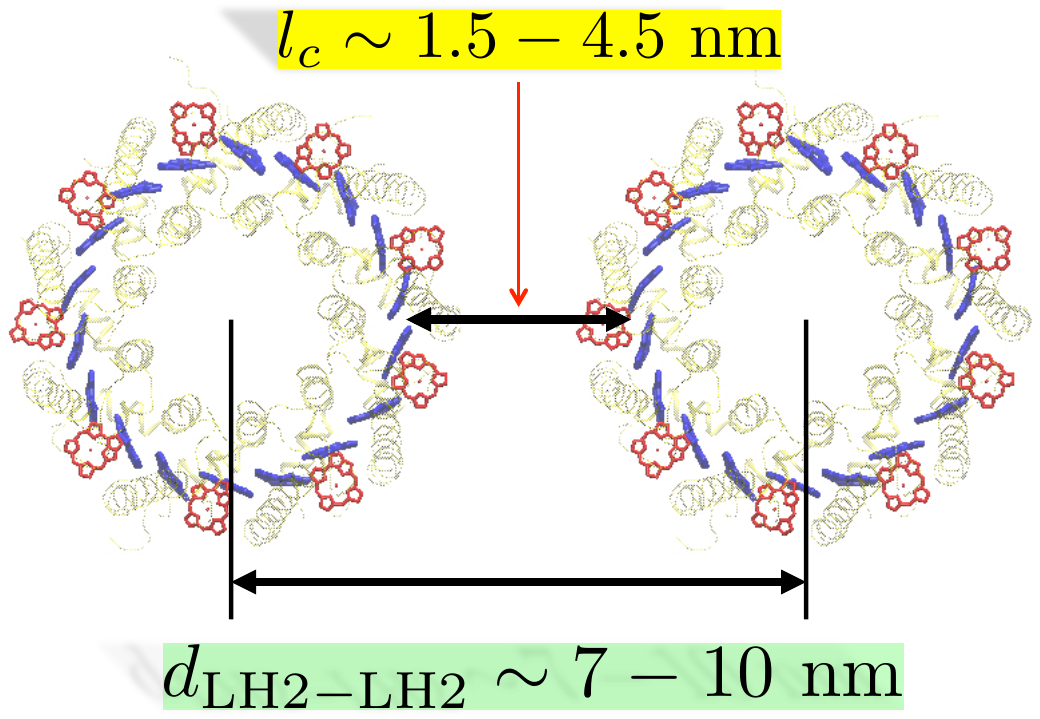} \vspace{-.5in}\\
\caption{The range of inter-LH2 distances estimated from AFM images.\cite{Scheuring-Science309,sturgis-biochemistry48,olsen-jbc283,Sumino-jpcb117} Note that $d_{\rm LH2-LH2}=l_c+2R_\beta$ (see Eq. (\ref{eq:r-beta-1})). Each LH2 complex here consists of 27 bactreiochlorophylls (BChls).  Nine of them facing up in the plane (red) form the B800 unit.  Eighteen of them showing sideways (blue) for the B850 unit. }
\label{fig:afm-dist}
\end{figure}

Since the pioneering computational work by Ritz {\it et al.},\cite{ritz-jpcb105} it has been widely \re{accepted}\cite{hu-qrb35,cogdell-qrb39,sundstrom-jpcb103,caycedo-soler-prl104} that the inter-LH2 exciton transfer takes about $10\ {\rm ps}$.  However, there is no experimental measurement firmly supporting this estimate yet.  \re{Low temperature} transient absorption spectroscopic data\cite{timpmann-jpcb104} indicate that inter-LH2 exciton transfer can take $1 - 7\ {\rm ps}$ or even longer.    Three-pulse photon echo peak shift measurement\cite{agarwal-jpcb105} suggests that  the average time of transfer is about ${\rm 5\ ps}$ at room temperature.   However, details of the arrangement of LH2 complexes in samples used for these spectroscopic studies are not clearly known. This makes quantitative modeling of such spectroscopic results a very difficult task.   The experimental data\cite{timpmann-jpcb104,agarwal-jpcb105} and the estimate by Ritz {\it et al.}\cite{ritz-jpcb105} for inter-LH2 exciton transfer time are roughly of the same order (${\rm ps}$ time scale).  Recent high level computational studies\cite{strumpfer-jcp134,strumpfer-jpcl3} also support such estimates.  However, these  do not yet serve as sufficient evidence for accepting ${\rm 10\ ps}$ as a solid estimate.  

\re{Let alone the quantitative information as noted above,  there are many conceptual questions that have significant implications. 
For example, which experimental conditions affect the long range exciton dynamics most?  Are there any physical variables that are difficult to control experimentally but have significant influence on how quantum effects contribute to inter-LH2 exciton dynamics?  What kinds of physical observables truly reflect quantumness of excitons and can be detected even in the presence of disorder? 
Addressing these questions is an important step 
for proper interpretation of experimental measurements as well. } 

Atomic force microscopy (AFM) data\cite{Scheuring-Science309,sturgis-biochemistry48,olsen-jbc283,Sumino-jpcb117} on patches of membrane show that the actual arrangement of LH2 complexes vary for different samples. In particular, the nearest-neighbor inter-LH2 distances appear to be quite variable.  Considering the sensitivity of rates on the distance, such a variation is expected to widen the distribution of inter-LH2 exciton transfer rates substantially. The actual arrangement of LH2 complexes of {\it in vivo} PSUs still remains unknown, but there is no reason to believe that they are more ordered than those observed from the AFM images.    Furthermore, even if there were no disorder in the structural arrangement of LH2 complexes, the inter-LH2 exciton dynamics depend significantly on the disorder in the excitation energy of each bacteriochlorophyll (BChl), the major pigment molecule in the LH2 complex.\cite{jang-jpcl6}   Thus, quantitative modeling of any spectroscopic signal relevant to inter-LH2 exciton transfer dynamics\cite{timpmann-jpcb104,agarwal-jpcb105} should involve averaging
over some ensemble, the identification of which also remains an important issue. 
      
In a recent work,\cite{jang-jpcl6} we have reported a computational study incorporating all-atomistic simulation, electronic structure calculation, and a coarse-grained quantum dynamics method called generalized master equation for modular exciton density (GME-MED),\cite{jang-prl113} and simulated \re{exciton dynamics between B850 units of LH2 complexes from {\it Rhodoblastus acidophilus}\cite{cogdell-qrb39} (formerly known as {\it Rhodopseudomonas acidophila})} for some select \re{set of parameters}.  This study\cite{jang-jpcl6} was motivated by the questions of whether and why the natural sizes of LH2 complexes, having only 8-10 fold symmetries, are optimal, a question that also motivated another theoretical study.\cite{cleary-pnas110}   Outcomes of our study\cite{jang-jpcl6} suggest that the optimality can be explained in terms of \re{maximum} stability of hydrogen bonding of BChls with 
tryptophan and tyrosine, \re{which results in} 
minimum energetic disorder and maximum inter-LH2 exciton transfer rates.  
The model developed in this work\cite{jang-jpcl6} and the GME-MED method\cite{jang-prl113}  can be employed to 
\re{examine how different characteristics of LH2 complexes influence the relay of } delocalized excitons, and to \re{gain} deeper insights into the interplay of quantum delocalization and the disorder for the realization of their functionality.  \re{An important first step to this end is} 
to 
\re{investigate} the dependence of inter-LH2 exciton kinetics on detailed molecular features, energetics, and the disorder as clearly as possible.  This work \re{addresses this issue through a comprehensive computational study of the exciton dynamics between two B850 units of LH2 complexes for {\it Rhodoblastus acidophilus}.} 
The model and approximations used in this work are similar to those of our previous work.\cite{jang-jpcl6}  
\re{Major features of the  model and methods  are  summarized below}.  

First, \re{an important assumption being made here}, as was done before,\cite{jang-jpcl6} is that the \re{the majority of the} inter-LH2 exciton dynamics can be represented by those between the B850 units, the lower bands of the LH2 complex where excitons reside predominantly.  \re{This assumption can be justified from the fact that the intra-LH2 transfer from its B800 unit to B850 unit is fast,\cite{jang-jpcb111} which takes about $1\ {\rm ps}$ on average. Further justification of this approximation will be provided later. } The current model for the B850 unit has evolved through previous computational studies and spectroscopic modeling.\cite{jang-jpcb105,jang-jcp118-2,jang-jpcb115,montemayor-jpcb122} 

The coordinates of BChls constituting the B850 unit, assuming that they are all on the same plane ($xy-$plane), can be expressed as   
\be
{\bf r}_{\alpha n}=R_\alpha \left ( \begin{array}{c}  \cos(2\pi n/9-\nu) \\
                                                                                            \sin(2\pi n/9-\nu)  \\
                                                                                            \end{array}
                                                                                    \right )\ , \label{eq:r-alpha-1}
 \ee             
 \be
{\bf r}_{\beta n}=R_\beta \left ( \begin{array}{c}  \cos(2\pi n/9+\nu) \\
                                                                                            \sin(2\pi n/9+\nu)  \\
                                                                                            \end{array}
                                                                                    \right )\ , \label{eq:r-beta-1}
 \ee 
\re{where the parameters were determined from molecular dynamics simulation trajectories as reported before,\cite{jang-jpcl6} and are respectively} $R_\alpha=2.604 \ {\rm nm}$, $R_\beta=2.753\ {\rm nm}$, $\nu=10.2^0$.  Note that the direction of $x$-axis is chosen to bisect the $\alpha_1$ and $\beta_1$ BChls and $2\nu$ is the angle between them.                                                                          
Assuming that the transition dipoles of $\alpha$ and $\beta$ BChls have the same axial angles, the three dimensional vectors of transition dipoles can be expressed as
\be
\mbox{\boldmath$\mu$}_{\alpha n}=\mu \left ( \begin{array}{c} \sin \theta \cos(2\pi n/9-\nu+\varphi_\alpha) \\
                                                                                            \sin \theta \sin(2\pi n/9-\nu+\varphi_\alpha)  \\
                                                                                            \cos\theta
                                                                                            \end{array} 
                                                                                    \right )\ ,  \label{eq:mu-alpha-1}
 \ee             
\be
\mbox{\boldmath$\mu$}_{\beta n}= \mu \left ( \begin{array}{c} \sin \theta \cos(2\pi n/9+\nu+\varphi_\beta) \\
                                                                                            \sin \theta \sin(2\pi n/9+\nu+\varphi_\beta)  \\
                                                                                            \cos\theta
                                                                                            \end{array} 
                                                                                    \right )\ , \label{eq:mu-beta-1}
 \ee    
 where $\theta=96.6^o$, $\varphi_\alpha=-106.6^o$, $\varphi_\beta=60.0^o$. The magnitude of the transition dipole moment, $\mu= 6.4{\rm\ D}$, was chosen based on previous studies.\cite{jang-jpcb111,jang-jpcl6,montemayor-jpcb122} 
 
The B850 unit of the second LH2 is represented by the following coordinates.    \be
{\bf r}'_{\alpha n}= \left ( \begin{array}{c}  R_\alpha \cos(2\pi n/9-\nu+\gamma) +d_{\rm LH2-LH2} \\
                                                                                            R_\alpha \sin(2\pi n/9-\nu+\gamma)  \\
                                                                                            \end{array}
                                                                                    \right )\ , \label{eq:r_alpha'}
 \ee             
 \be
{\bf r}'_{\beta n}=\left ( \begin{array}{c}  R_\beta \cos(2\pi n/9+\nu+\gamma) +d_{\rm LH2-LH2}\\
                                                                                          R_\beta  \sin(2\pi n/9+\nu+\gamma)  \\
                                                                                            \end{array}
                                                                                    \right ) \ ,  \label{eq:r_beta'}
 \ee 
 where $d_{\rm LH2-LH2}$ is the center-to-center distance of the two LH2 complexes and $\gamma$ is the in-plane rotation angle of the second LH2 complex. 
Similarly, the coordinates of the transition dipoles of the 2nd B850 can be expressed as
\be
\mbox{\boldmath$\mu$}_{\alpha n}'=\mu \left ( \begin{array}{c} \sin \theta \cos(2\pi n/9-\nu+\varphi_\alpha+\gamma) \\
                                                                                            \sin \theta \sin(2\pi n/9-\nu+\varphi_\alpha+\gamma)  \\
                                                                                            \cos\theta
                                                                                            \end{array}
                                                                                    \right )  \ ,
 \ee             
\be
\mbox{\boldmath$\mu$}_{\beta n}'=\mu \left ( \begin{array}{c} \sin \theta \cos(2\pi n/9+\nu+\varphi_\alpha+\gamma) \\
                                                                                            \sin \theta \sin(2\pi n/9+\nu+\varphi_\alpha+\gamma)  \\
                                                                                            \cos\theta
                                                                                            \end{array}
                                                                                    \right ) \ .
 \ee

The total exciton-bath Hamiltonian for the pair of B850 units can be expressed as
\be\label{E1}
H_T = \epsilon_g|g\rangle\langle g|+H_{e,1}^0+H_{e,2}^0+H_{e,c}+\delta H_{e} + H_b + H_{eb} \ , 
\ee
where $| g \rangle$ is the ground electronic state with energy $\epsilon_g$, $H_{e,k}^0$ is the exciton Hamiltonian for the $k$the B850 unit \re{in the absence of disorder}, $H_{e,c}$ is the electronic coupling Hamiltonian between the two B850 units, and $\delta H_e$ represents the disorder in the exciton Hamiltonian.  The detailed expression for $H_{e,k}^0$ is as follows.
\ben
H_{e,k}^0&=&\sum_{n=1}^9  \left \{\epsilon_{\alpha_k} |\alpha_{k,n}\rangle \langle \alpha_{k,n}|+\epsilon_{\beta_k}  |\beta_{k,n}\rangle \langle \beta_{k,n}| \right . \nonumber \\
&&+\left . J_{\alpha \beta}(0) (|\alpha_{k,n}\rangle \langle \beta_{k,n}|+|\beta_{k,n}\rangle \langle \alpha_{k,n}| )  \right\} \nonumber \\
&+&\sum_{n=1}^9 \sum_{m\neq n}^9\sum_{s,q=\alpha}^\beta J_{sq}(n-m)|s_{k,n}\rangle \langle q_{k,m}| \ ,  \label{eq:he0}
\een
where $|\alpha_{k,n}\rangle$ ($|\beta_{k,n}\rangle$) represents the excited state where only $\alpha_{k,n}$-BChl ($\beta_{k,n}$-BChl) is excited, with the excitation energy $\epsilon_{\alpha_k}$ ($\epsilon_{\beta_k}$), and $J_{sq}(n-m)$ is the electronic coupling between $|s_{k,n}\rangle$ and $|q_{k,m}\rangle$ states.   The two nearest neighbor electronic couplings determined before,\cite{montemayor-jpcb122} $J_{\alpha\beta}(0)=239 \ {\rm cm^{-1}}$ and $J_{\alpha\beta}(1)=140\ {\rm cm^{-1}}$, are used here.  For all other non-neighbor electronic couplings within each B850 unit, the following transition dipole interactions are used.
\be
J_{sq}(n-m)=\frac{C\kappa_{sn,qm}}{|{\bf d}_{sn,qm}|^3}\ , \label{eq:td-int}
\ee  
where ${\bf d}_{sn,qm}={\bf r}_{sn}-{\bf r}_{qm}$, $C=223.43578$ (in the units where the distances are in ${\rm nm}$ and the electronic couplings are in ${\rm cm^{-1}}$), and $\kappa_{sn,qm}$ is the following orientational factor for the transition dipole interaction: $
\kappa_{sn,qm}=\mbox{\boldmath$\mu$}_{s n}\cdot\mbox{\boldmath$\mu$}_{q m}/\mu^2 - 3(\mbox{\boldmath$\mu$}_{s n}\cdot {\bf d}_{sn,qm})(\mbox{\boldmath$\mu$}_{q m}\cdot {\bf d}_{sn,qm}/(\mu^2|{\bf d}_{sn,qm}|^2)$.  \re{Note that the value of $C$ above also includes the screening effect within the LH2 complex.}
Likewise, the electronic coupling Hamiltonian between two B850 units is also given by
\be
H_{e,c}=\sum_{s,q=\alpha}^\beta\sum_{n,m=1}^9J_{sn,qm}^{c} (|s_{1,n}\rangle\langle q_{2,m}|+|q_{2,m}\rangle\langle s_{1,n}|)\ ,
\ee
where
\be
J_{sn,qm}^{c}=\frac{C\kappa_{sn,qm}^{c}}{|{\bf d}_{sn,qm}^c |^3}\ ,
\ee
with  ${\bf d}_{sn,qm}^c = {\bf r}_{sn}-{\bf r}'_{qm}$ and $\kappa_{sn,qm}^{c}=\mbox{\boldmath$\mu$}_{s n}\cdot\mbox{\boldmath$\mu$}'_{q m}/\mu^2 - 3(\mbox{\boldmath$\mu$}_{s n}\cdot  {\bf d}_{sn,qm}^c )(\mbox{\boldmath$\mu$}'_{q m}\cdot {\bf d}_{sn,qm}^c)/(\mu^2|{\bf d}_{sn,qm}^c|^2)$.  \re{The use of $C$ above amounts to assuming that the screening factor between two different LH2 complexes is the same as that within each LH2 complex.  While this assumption may incur some errors considering the distance dependence of screening factors in general, they are not expected to alter major conclusions of this work.} 

The term $\delta H_e$ in Eq. (\ref{E1}), which represents the disorder, is given by
\be
\delta H_e= \sum_{k=1}^2\sum_{n=1}^9  \left \{ \delta \epsilon_{\alpha_k,n} |\alpha_{k,n}\rangle\langle \alpha_{k,n}|+\delta \epsilon_{\beta_{k,n}}|\beta_{k,n}\rangle \langle \beta_{k,n}| \right\}  \ , \label{eq:delhe}
\ee
where $\delta \epsilon_{\alpha_{k,n}}$ and $\delta \epsilon_{\beta_{k,n}}$ are specific realizations of the disorder in $\epsilon_{\alpha_k}$ and $\epsilon_{\beta_k}$.  
$H_b$ in Eq. (\ref{E1}) represents the bath, all the vibrational and environmental degrees of freedom, and  is modeled as an independent sum of harmonic oscillators as follows:
$H_b = \sum_j \hbar \omega_j ( b^\dagger_{j} b_{j} + 1/2 )$,  
where $b^\dagger_{j} (b_{j})$ is the creation (annihilation) operator of the $j$th harmonic oscillator.  All the bath modes are collectively expressed by a single index $j$. 
The last term in Eq. (\ref{E1}) represents the exciton-bath interaction Hamiltonian and is assumed to have the following form: 
$H_{eb} = \sum_{k=1}^2\sum_j \sum_{n=1}^9 \sum_{s=\alpha}^{\beta} \hbar g_{sn,j} \big( b_{j} + b^\dagger_{j} \big) | s_{k,n} \rangle \langle s_{k,n} |$, 
where $g_{sn,j}$ represents the exciton-bath coupling of the $j$th mode to the site excitation state $|s_{k,n}\rangle$.   \re{The definition of the spectral density and the model of the spectral density being used in this work is provided in the Supporting Information (SI).}

Starting from an initial condition where the exciton is in thermal equilibrium in the first B850 unit (denoted as 1), the ensuing time dependence of its exciton population was calculated in the presence of the second B850 unit employing the following time-local form of GME-MED:\cite{jang-prl113}
\be
\frac{\partial}{\partial t} p_1(t) =\left  \{W_{2 \rightarrow 1}(t) p_2(t)-W_{1 \rightarrow 2}(t)  p_1(t)\right \}\ ,\label{eq:master_2}
\ee 
where  $p_1(t)$ and $p_2(t)$ are probabilities to find excitons in the two B850 units, respectively.  \re{The expression for $W_{2\rightarrow 1}(t)$ ($W_{1\rightarrow 2}(t)$) and the approximations involved are detailed in the SI.} 
\begin{widetext}

\begin{figure}
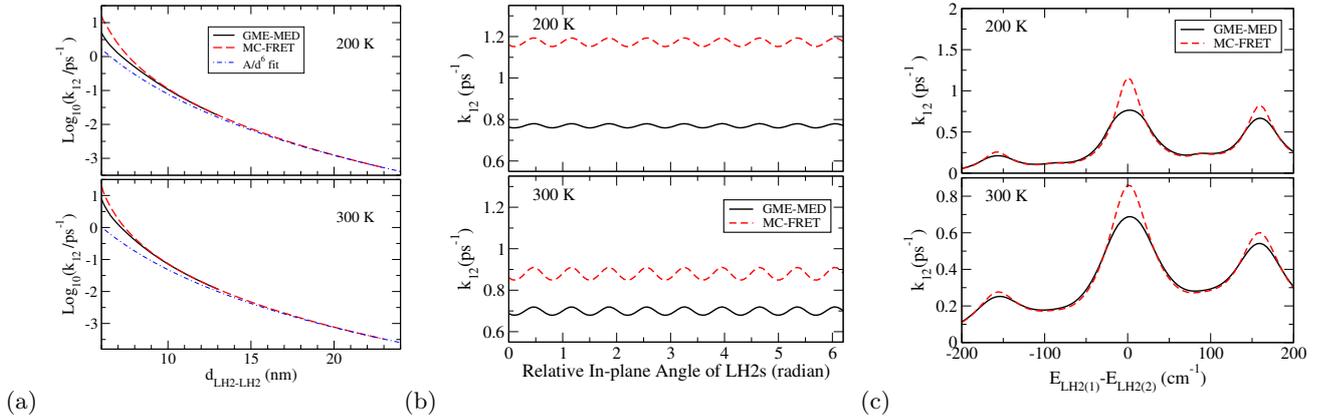

(a)\includegraphics[width=1.8in]{Jang_Figure2a.eps}\hspace{.1in}(b)\includegraphics[width=2.05in]{Jang_Figure2b.eps} \hspace{.1in}(c)\includegraphics[width=2.1in]{Jang_Figure2c.eps}   
\caption{(a) Dependence of GME-MED and MC-FRET rates on the LH2-LH2 distances. The value of $A$ for T=200 K is 78190 and that for T=300 K is 48215. (b) Dependence of rates on relative angles between two B850 complexes at T= 200 K (upper panel)  and 300 K (lower panel) with $l_c=2\ {\rm nm}$, for which $d_{LH2-LH2}=7.506\ {\rm nm}$.  The GME-MED rates and the MC-FRET rates are compared for each case.  (c) Dependence of rates on relative energies between two B850 complexes at T= 200 K (upper panel)  and 300 K (lower panel) with $l_c=2\ {\rm nm}$, for which $d_{LH2-LH2}=7.506\ {\rm nm}$.  The GME-MED rates and the MC-FRET rates are compared for each case.}
\label{fig:d_erate}
\end{figure}

\end{widetext}

For the determination of the effective forward rate from the population dynamics, a transfer time $\tau_{tr}$ can be defined as the shortest time satisfying  the following relation:
\be
\ln \left [p_1(\tau_{tr})-\frac{Z_1}{Z_2} p_2(\tau_{tr}) \right ]=-1 \ . \label{eq:tau_tr}
\ee
Then, an effective forward rate can be defined as 
\be 
k_f=\frac{Z_2}{Z_1+Z_2}\frac{1}{\tau_{tr}} \ , \label{eq:kf}
\ee
where $Z_k=\sum_{p_k} e^{-\tilde \epsilon_{p_k}/k_BT}$, the exciton partition function of the B850 unit of the $k$th LH2.  \re{It is important to note that this includes the diagonal contributions of the exciton-bath coupling, and thus accounts for  the contribution of the reorganization energy to the equilibrium population.}

The effective forward rate defined by Eq. (\ref{eq:kf}) \re{satisfies} the detailed balance and \re{amounts to assuming that the population relaxation rate  is equal to} the inverse of $\tau_{tr}$, the time when the population difference becomes an $1/e$ \re{factor} of its initial value.  For genuine rate processes, this agrees with the actual forward rate.  For more general case but with monotonic population decay, this can serves as a good measure for the effective forward rate. For the remainder of this work, we thus term Eq. (\ref{eq:kf}) as the effective GME-MED rate.  \re{Examples of actual population dynamics when the GME-MED rate is different from MC-FRET rate are shown in the SI.}  \\

\noindent
{\it Dependence of rates on $d_{\rm LH2-LH2}$}: Figure 2(a) shows distance dependences of the GME-MED rates and the MC-FRET rates at two different values of temperature, 200 K and 300 K.   Also shown are fitting curves of these data in the asymptotic limit by a function $A/d_{\rm LH2-LH2}^6$, which represents the FRET rate\cite{forster-ap,forster-dfs} for transition dipole interactions.  For the values of $d_{\rm LH2-LH2}$ in the range of $5-25\ {\rm nm}$ that are being considered here, the rates change by about four orders of magnitudes as is clear from the logarithmic scale in Fig. \ref{fig:d_erate}(a).   Deviation from the FRET-like behavior with transition-dipole interaction becomes apparent for distances smaller than ${\rm 15\ nm}$, about three times the diameter of LH2, where the rates are as small as $\sim 0.01\ {\rm ps^{-1}}$.

It is important to note that both the GME-MED rate and the MC-FRET rate exhibit much steeper distance dependences than the FRET rate.  This pattern clearly demonstrates that the combination of multipolar effects and the dark exciton states make positive contributions to \re{the transfer of excitons}, and \re{their magnitudes become substantial as the distance becomes shorter}.  For example, for $d_{\rm LH2-LH2}= {\rm 6 \ nm}$, the factor of enhancement turns out to be about $10$.

 In general, contribution of multipolar effects does not always lead to the increase of exciton transfer rate.\cite{scholes-arpc54,fuckel-jcp128,yang-jacs132}  For some types of chromophores and relative configurations of the donor and the acceptor, they can in fact result in reduction of the rate.  In this sense, the level of enhancement for the inter-LH2 exciton transfer rate, as observed in Fig. \ref{fig:d_erate}(a), reflects significant quantum effects.   This indicates that the quantum delocalization does more than simply increasing effective dipole moment because if so the distance dependence should remain the same.   Thus, the distance dependence of inter-LH2 exciton transfer demonstrated here serves as a good example that the multipolar effects and contributions from dark exciton states, both  originating from the delocalized quantum nature of excitons, provide a robust mechanism  to make substantial enhancement of the exciton transfer rate.

For the particular case of two degenerate LH2s without disorder as considered above, the GME-MED rates are smaller than the MC-FRET rates.  In other words, non-Markovian effects due to finite time scale of intra-LH2 exciton dynamics cause the rate to be smaller than the prediction of MC-FRET.  This may not always be the case for other situations. At least, it is clear from Fig. \ref{fig:d_erate}(a) that such non-Markovian effects become more significant as the distance becomes shorter.  However, the discrepancy between the two is in fact rather small in the range of physical distances observed from AFM images, ${\rm 7 - 10\ nm}$ (see Fig. \ref{fig:afm-dist}), and  at $T={\rm 300\ K}$.   This suggests that the use of MC-FRET is reasonable for modeling the exciton dynamics in natural aggregates of LH2 complexes at room temperature, even without considering the disorder.  As will be seen below, inclusion of the disorder makes the discrepancy between the two even smaller.
 
\noindent
{\it Dependence of rates on $\gamma$}: Figure 2(b) shows the dependences of rates on the angle $\gamma$, the relative in-plane rotational angle of the two B850 units (see Eqs. (\ref{eq:r_alpha'}) and (\ref{eq:r_beta'})).   The periodic modulation of the rate reflects the nine-fold symmetry of the LH2 complex.     The variation with angle is slightly larger for T=300 K than for T=200 K.  This shows that more delocalization of exciton in the lower temperature makes the rate less sensitive to the angle.  \re{Note that this effect cannot be explained in terms of conventional temperature dependence of the contribution of the bath spectral density, which is independent of the rotational angle $\gamma$. }  Overall, the variation of the rate with the angle $\gamma$ is fairly small.  The range of modulation is less than 10\% of the mean value.   These are even smaller than the discrepancies between GME-MED and MC-FRET rates.  Thus, one can view that the inter-LH2 exciton transfer rate is virtually isotropic.   This is even more true \re{for the rate averaged over the ensemble of disorder as will be shown later.}\\

\begin{widetext}

\begin{figure}
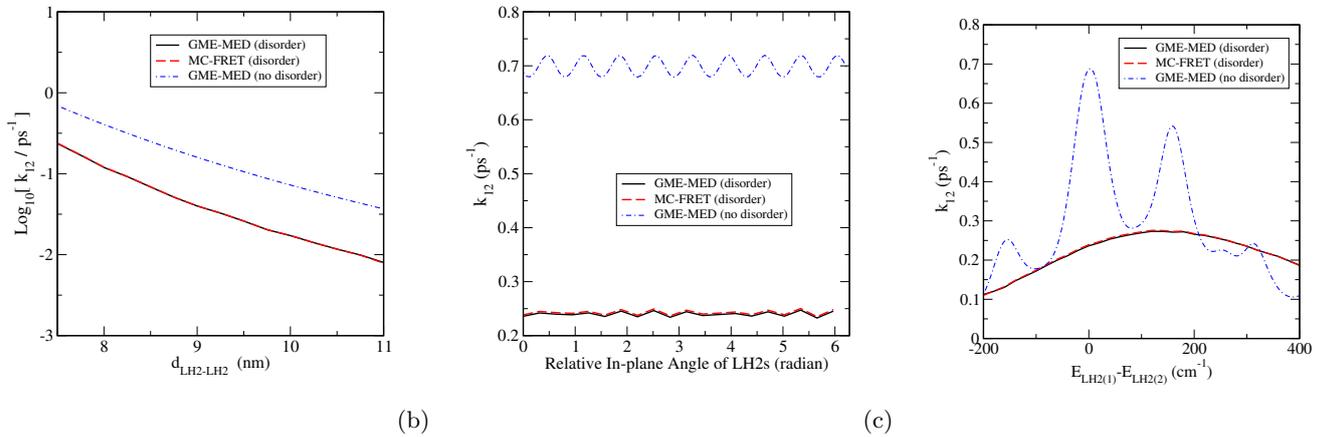

(a)\includegraphics[width=2.in]{Jang_Figure3a.eps}\hspace{.2in}
(b)\includegraphics[width=2.in]{Jang_Figure3b.eps}\hspace{.2in}
(c)\includegraphics[width=2.in]{Jang_Figure3c.eps}
\caption{Dependences of average values of GME-MED rates and MC-FRET rates at ${\rm 300\ K}$.  Each point was determined by averaging over an ensemble of 10,000 realizations of the disorder in site excitation energies of BChls with standard deviation, $200\ {\rm cm^{-1}}$.  (a) The distance dependence for $\gamma=0$ and $E_{LH2(1)}-E_{LH2(2)}=0$.  (b) The dependence on relative angle for $d_{LH2-LH2}=7.506\ {\rm nm}$ and $E_{LH2(1)}-E_{LH2(2)}=0$. (c) The dependence on $E_{LH2(1)}-E_{LH2(2)}$,  the bias of average excitation energies between two LH2 complexes, for $d_{LH2-LH2}=7.506\ {\rm nm}$ and $\gamma=0$. }
\end{figure}

\end{widetext}
\noindent
{\it Dependence of rates on \re{relative energies of LH2 complexes}}: Figure 2(c) shows the \re{variation} of exciton transfer rates with the \re{change of relative excitation energies of two LH2 complexes}, $E_{LH2(1)}-E_{LH2(2)}=\epsilon_{\alpha_1}-\epsilon_{\alpha_2}$.  \re{Here, note that $\epsilon_{\alpha_1}$ and $\epsilon_{\alpha_2}$ refer to the excitation energies $\alpha$ BChl entering the  Hamiltonian in the absence of disorder, $H_{e,k}^0$ of Eq. (\ref{eq:he0}), which are assumed to be the same as those for $\beta$ BChls.  These also correspond to average excitation energies of BChls in the presence of disorder.}    The pronounced rates near $0$, $150$, and $-150\  {\rm cm^{-1}}$ reflect the discrete quantum nature of exciton levels and the broadening due to exciton-bath coupling within each B850 unit.  The resonance becomes significant when there is matching of exciton states, which results in faster rates.  It is interesting  to note that the MC-FRET rate overestimates this enhancement when compared with the GME-MED rate. 

The variation of the rate with the \re{relative LH2 energy} is much larger than that due to the in-plane rotation angle $\gamma$.  As yet, its extent is relatively moderate within the range of energies considered.  The maximum factor of difference of the transfer rates within the range considered is about 7.  The range of  actual $E_{LH2(1)}-E_{LH2(2)}$ in aggregates of LH2 complexes is expected to be less than $100\ {\rm cm^{-1}}$ according to modeling of spectroscopic data.\cite{jang-jpcb115,montemayor-jpcb122}  Thus, the actual variation of the rate due to the energy difference is expected to be even smaller.  If the exciton transfer is only between those due to single pigment molecules, the transfer rate is likely to be much more sensitive to the relative energy.  

\begin{figure}
\includegraphics[width=2.7in]{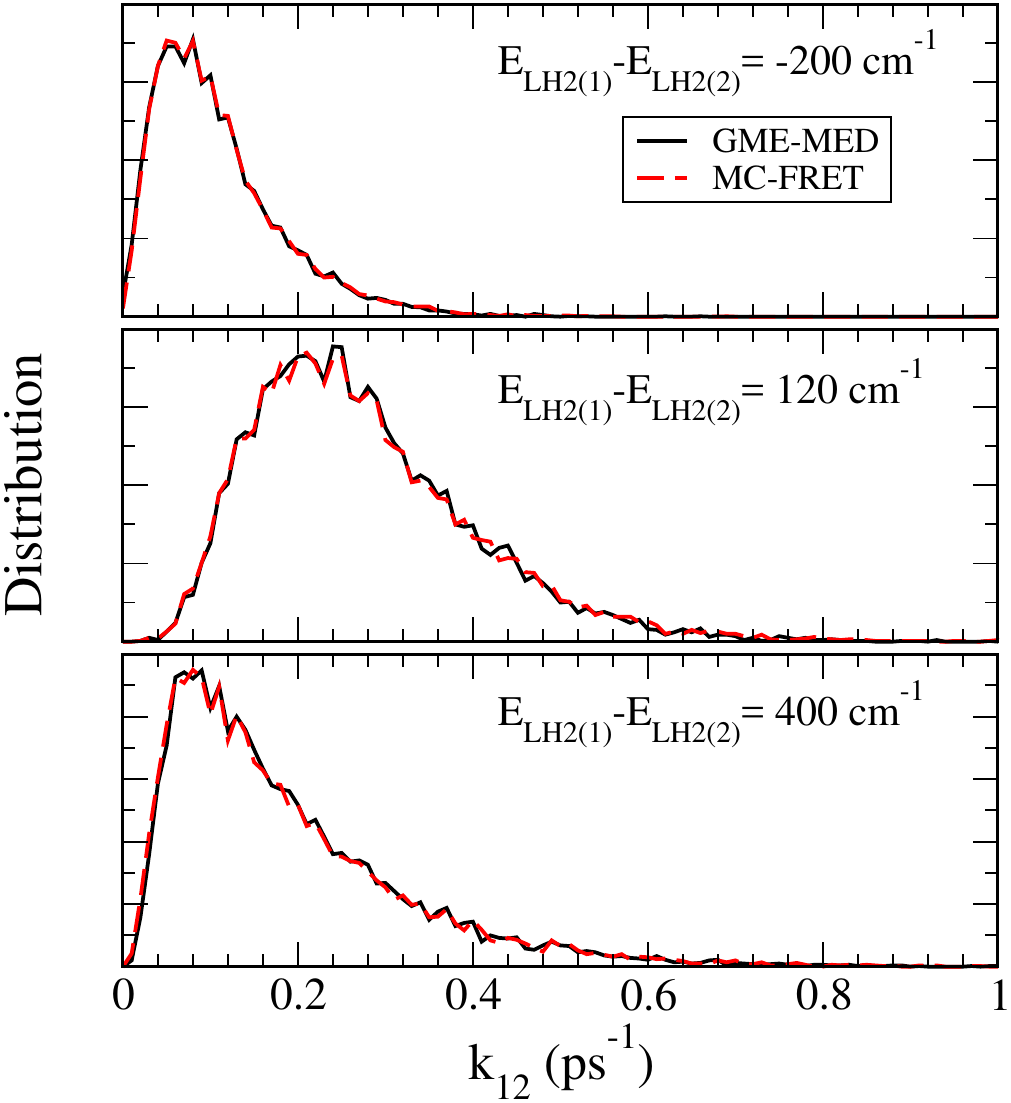}
\caption{Distributions of inter-LH2 transfer rates for $d_{LH2-LH2}=7.506\ {\rm nm}$ and $\gamma=0$ at ${\rm 300\ K}$.}
\end{figure}

\noindent
{\it Dependence of rates on the disorder}: In order to understand the effects of the disorder on the inter-LH2 exciton transfer rate, GME-MED calculations were conduced for each ensemble of 10,000 realizations of the Gaussian disorder in excitation energies of BChls, with the standard deviation $\sigma=200\ {\rm cm^{-1}}$, and  at $T=300\ {\rm K}$.  Figure 3 shows the resulting average rates with the variation of inter-LH2 distance, angle, and the relative energy.  The corresponding GME-MED results without the disorder, which were shown in Fig. 2, are also compared here.  

Figure 3(a) shows that the disorder reduces the average rate by about a factor of $4$ and makes the distance dependence slightly steeper.    However, the qualitative nature of the distance dependence does not change significantly.   In other words, the quantum \re{characteristics} of the distance dependence are robust against the disorder.  Table \ref{table:rate-dist} provides numerical values of average GME-MED rates and their standard deviations  for seven values of inter-LH2 distances.  
These show that the average transfer time of $10\ {\rm ps}$ estimated by Ritz {\it et al.}\cite{ritz-jpcb105} corresponds only to $d_{\rm LH2-LH2}=8 \sim 8.25{\rm\ nm}$.   On the other hand, for $d_{\rm LH2-LH2}=7.5 \sim 9{\rm\ nm}$, the range of inter-LH2 transfer time is estimated to be about $4 \sim 25\ {\rm ps}$.

Figure 3(b) shows that the disorder almost removes the variation of the inter-LH2 exciton transfer rate with the angle $\gamma$.  There is practically no difference between the average GME-MED and MC-FRET rates.   Figure 3(c) shows that the disorder removes the resonance structure and makes the energy dependence similar to that of classical Marcus-like\cite{marcus-jcp24,marcus-jcp43,marcus-bba811} inverted parabola but with much weaker reorganization energy.  The averaged GME-MED and MC-FRET rates almost agree here as well.   In another words, the disorder has an effect of washing out non-Markovian effects.   It is also important to note that the dependence of the average rate on the \re{relative energy of LH2s} has become much more moderate than that without disorder.  Considering that the \re{standard deviation of the disorder in the excitation energy of LH2 as a whole (not that of the individual BChl)} is less than $100\ {\rm cm^{-1}}$ according to spectroscopic modeling,\cite{jang-jpcb115,montemayor-jpcb122} \re{ the corresponding average rate is expected to vary less than 20\% of that between two LH2s with the same energy}.  In other words, the combination of quantum effects and the disorder in this case make the inter-LH2 transfer rate robust against the relative energy difference of LH2s.  

\re{Figure 4 shows distributions of MC-FRET and GME-MED rates for three different values of relative LH2 energies for $d_{LH2-LH2}=7.506 \ {\rm nm}$. It is clear that there is no appreciable discrepancy even at the level of distributions between MC-FRET and GME-MED rates.  Examples of average population dynamics shown in the SI also confirm this fact.  Thus, the disorder seems to genuinely wash out the non-Markovian effects and make MC-FRET rate  as reliable alternative to full GME-MED calculation for inter-LH2 exciton transfer dynamics.  } 

 \begin{table} 
\caption{Average GME-MED rates and their standard deviations for different values of the inter-LH2 distance $d_{\rm LH2-LH2}$, calculated for an ensemble of 10,000 realizations of the Gaussian disorder in the excitation energies of BChls with standard deviation $200\ {\rm cm^{-1}}$ at 300 K.  }
\begin{tabular}{ccc}
\hline
\hline
$d_{\rm LH2-LH2}$ (${\rm nm}$) & $\langle k_{1\rightarrow 2} \rangle $ (${\rm ps^{-1}}$) & Stand. dev. of  $ k_{1\rightarrow 2}$'s (${\rm ps^{-1}}$) \\
\hline
$7.506$ & $0.234$  & $0.111$\\
$7.756$ & $0.168$  & $0.0809$\\ 
$8.006$ & $0.119$   & $0.0552$ \\
$8.256$ & $0.0915$ & $0.0408$\\
$8.506$ & $0.0681$ & $0.0325$ \\
$8.756$ & $0.0510$ & $0.0235$\\
$9.006$ & $0.0397$ & $0.0183$\\
\hline
\hline
\end{tabular}
\label{table:rate-dist}
\end{table}

In summary, this work reports comprehensive computational results elucidating the dependences of inter-LH2 exciton transfer rates, \re{modeled as those between B850 units}, on three major physical features,  and also the effects of disorder on such dependences.   These data are the first of this kind 
and will serve as valuable resources for quantitative understanding of the trends on exciton transfer kinetics between LH2 complexes.  
Most importantly, these results offer \re{new and important}  information on how quantum delocalization of excitons, non-Markovian effects, and the disorder contribute to the exciton transfer kinetics, as \re{highlighted}  below. 

\begin{itemize}
\item Quantum delocalization of excitons results in significant deviation \re{in the distance dependence of the exciton transfer rate} from the trend \re{expected} for \re{that between} transition dipoles.  \re{The deviation is appreciable} 
at distances shorter than three times the diameter of the LH2 complex, and \re{reaches up to a factor of 10 at the closest distance tested}.  This also means that the inter-LH2 exciton transfer rates for \re{the nearest and the second nearest neighbors of LH2 complexes cannot be modeled accurately} by conventional FRET rate for transition dipole interactions. 

\item  \re{The fact that the exciton-transfer rate is almost isotropic in the absence of disorder except for minor modulation, can only be explained in terms of delocalization of excitons.    In the presence of disorder,  the rate for each realization of disorder can be less isotropic.  However, the average rate for the entire ensemble of the disorder is even more isotropic because the ensemble of disorder does not have specific preference for certain direction.}   

\item The inter-LH2 exciton transfer rates show resonance structure with respect to the relative energies of LH2 complexes, which however disappears once averaging over the disorder is taken.

\item Within typical inter-LH2 distances observed from AFM images, the non-Markovian effects due to finite time scale of intra-LH2 exciton dynamics do not make significant contributions to the inter-LH2 exciton transfer rate.  Such effects become even less significant when averaged over the disorder.

\end{itemize}  

\re{As stated above,} the qualitative trend of the distance dependence and near isotropic nature of the inter-LH2 exciton transfer rates, both of which originate from delocalized nature of excitons,  \re{appear to be intact on average even in the presence of the disorder}.  On the other hand, the non-Markovian effect, subtle modulation of inter-LH2 rate with respect to the angle, and the resonance structure with respect to the relative excitation energies of LH2 are \re{found to be fragile} when disorder is introduced.    \re{Of course, the fragility refers to the specific features of ensemble averaged physical observables, not necessarily the quantumness of exciton transfer mechanism at individual pair level. } 

\re{The effect of the disorder on the energy dependence of rates}, as demonstrated by Fig. 3(c), has significant implications in broader context.  
The near inverted parabolic dependence of the average rate with respect to the \re{relative LH2 energies} could easily be \re{misinterpreted} as an evidence for classical rate behavior with very small reorganization energy.  However, the analysis demonstrated here clarifies that such apparent classical-like behavior is simply due to the fragility of the resonance structure \re{that appears in} Fig. 2(c).    

\re{There are two important issues that require some clarification and further investigation.   First, the contribution of the B800 unit needs to be examined more thoroughly.  Transfer rates between B850 units are much slower than those for the intra-LH2 transfer from B800 to B850, which take about 1 ps.  Since the energy difference between the exciton energies of two units is significantly larger than the thermal energy at room temperature, it is reasonable to assume that the B800 to B850 transfer can be viewed almost as irreversible and the majority of the exciton population (about 95\% or more) resides in the B850 unit. Therefore, even though direct inter-LH2 transfer through B800 units were possible because of their proximity, their net contribution to inter-LH2 exciton dynamics is expected to be marginal.  Still, considering the dispersive nature of the dynamics and the possibility of some kind of correlated transfer, it is worthwhile to examine the contribution of the B800 unit more carefully.   Recently completed models for the full LH2\cite{montemayor-jpcb122} and the GME-MED method\cite{jang-prl113} in fact make such investigation straightforward, which will be conducted in the near future.} 

\re{Second, the approximations involved in the current implementation of the GME-MED method\cite{jang-prl113} need to be tested against more accurate methods.  The distance corresponding to inter-LH2 transfer time of 10 ps is about 8.1 nm according to Table I.  This is similar to the estimate for the distance (about 8 nm) based on the information available from the work\cite{ritz-jpcb105,strumpfer-jcp134} by Schulten's group.  Therefore, the two appear to be consistent.   Still, it would be much more satisfactory to compare the GME-MED method with more accurate methods for exactly the same model and parameters.  This will be the subject of future investigation as well. } 

Some quantitative details   presented in this work may need to be modified  once more accurate quantum dynamics calculations are conducted and more refined models are employed.  In addition, the effects of time dependent fluctuations that cannot be represented by quantum bath or static disorder need to be understood better.  However, considering the reasonability of assumptions and models employed here, it is expected that such modification is not likely to alter the major conclusion regarding 
      \re{the quantum effects and their interplay with the disorder}. 
\ \vspace{2in}\\

\acknowledgements
This work was mainly supported by the Office of Basic Energy Sciences, Department of Energy (DE-SC0001393), and partially supported by the National Science Foundation  (CHE-1362926) for the development of quantum dynamics methods. The author also acknowledges computational support by the Queens College Center for Computational Infrastructure for the Sciences.

\providecommand{\latin}[1]{#1}
\providecommand*\mcitethebibliography{\thebibliography}
\csname @ifundefined\endcsname{endmcitethebibliography}
  {\let\endmcitethebibliography\endthebibliography}{}

 \ \vspace{5in}\\

\newpage
\newpage

\renewcommand{\theequation}{S\arabic{equation}}
\renewcommand{\thefigure}{S\arabic{figure}}
\renewcommand{\thetable}{S\arabic{table}}
\renewcommand\thepage{S\arabic{page}}
\setcounter{page}{1}
\setcounter{figure}{0}
\setcounter{equation}{0}
\begin{widetext}
\appendix
\section{Supporting Information: Robust and Fragile Quantum Effects in the Transfer Kinetics of Delocalized Excitons between LH2 Complexes}

\subsection{Spectral Density and Lineshape functions}
 It is assumed that there is only one local contribution to the spectral density, which is defined  as follows: 
\be\label{E17}
{\mathcal J}(\omega) = \pi\hbar \sum_j g^2_{\alpha(\beta) n, j} \delta(\omega-\omega_j) \ .
\ee
For the above spectral density, the following model that has been used in previous works\cite{jang-jpcb111,jang-jpcl6,montemayor-jpcb122} is used.
\ben
{\mathcal J}(\omega)&=&\pi\hbar \Big (\eta_1 \omega e^{-\omega/\omega_{c1}}+\eta_2\frac{\omega^2}{\omega_{c2}}e^{-\omega/\omega_{c2}} +\eta_3\frac{\omega^3}{\omega_{c3}^2}e^{-\omega/\omega_{c3}} \Big)\ , \label{eq:spd-lh2}
\een
where $\eta_1=0.22$, $\eta_2=0.78$, $\eta_3=0.31$, $\hbar\omega_{c1}=170\ {\rm cm^{-1}}$, $\hbar\omega_{c2}=34\ {\rm cm^{-1}}$, and $\hbar\omega_{c3}=69\ {\rm cm^{-1}}$.

In applying the GME-MED method,\cite{jang-prl113} it is necessary to calculate line shape functions and reorganization energy as defined below.  
\ben
&&\lambda =\frac{1}{\pi} \int_0^\infty d\omega  \frac{{\mathcal J}(\omega)}{\omega}  \ ,  \label{eq:lambda} \\
&&G_{i}(t)=\frac{1}{\pi\hbar}\int_0^\infty d\omega \frac{{\mathcal J}(\omega)}{\omega^2} \sin(\omega t) \ , \label{eq:gi}\\
&&G_{r}(t)= \frac{1}{\pi \hbar}\int_0^\infty d\omega \frac{{\mathcal J}(\omega)}{\omega^2} \coth\left (\frac{\hbar \omega}{2k_BT}\right )(1-\cos (\omega t)) \  ,  \label{eq:gr} 
\een
where $k_B$ is the Boltzman constant.
For the spectral density of Eq. (\ref{eq:spd-lh2}),  the exact expressions for $\lambda$ and $G_i(t)$ are as follows:
\ben
&&\lambda=\hbar (\eta_1\omega_{c,1}+\eta_2\omega_{c,2}+2\eta_3\omega_{c,3})\ , \\
&&G_i(t)= \eta_1 \tan^{-1}(\tau_{1,0})+\eta_2 \frac{\tau_{2,0}}{1+\tau_{2,0}^2}\nonumber \\
&&\hspace{.4in} +2\eta_3 \frac{\tau_{3,0}}{(1+\tau_{3,0}^2)^2}\ ,
\een
where $\tau_{i,0}=t\omega_{c,i} $.  On the other hand, for $G_r(t)$, an interpolation formula\cite{jang-jpcb106} for $\coth(\hbar\omega/2k_BT)$ leads to the following approximation:
\ben
&&G_r(t)\approx \frac{\eta_1}{2}\Big \{ \ln (1+\tau_0^2)+2\ln (1+\tau_1^2)+2\ln(1+\tau_2^2) .\nonumber \\
&&\hspace{.6in} +4\frac{(1+5\theta_1/2)}{\theta_1}\int_0^{\tau_{5/2}} d\tau' \tan^{-1} (\tau') \Big\} \nonumber \\
&&\hspace{.4in}+\eta_2 \Big \{ \frac{\tau_0^2}{1+\tau_0^2}+\frac{2}{(1+\theta)}\frac{\tau_1^2}{1+\tau_1^2}\nonumber \\
&&\hspace{.6in}+\frac{2}{(1+2\theta)}\frac{\tau_2^2}{(1+\tau_2^2)}+\frac{1}{\theta}\ln (1+\tau_{5/2}^2 )\Big\} \nonumber\\
&&\hspace{.4in}+2\eta_3 \Big \{ \frac{\tau_0^4+3\tau_0^2}{(1+\tau_0^2)^2}+\frac{2}{(1+\theta)^2}\frac{\tau_1^4+3\tau_1^2}{(1+\tau_1^2)^2}\nonumber \\
&&\hspace{.6in}+\frac{2}{(1+2\theta)^2}\frac{\tau_2^4+3\tau_2^2}{(1+\tau_2^2)^2}\nonumber \\
&&\hspace{.6in}+\frac{2}{\theta(1+5\theta/2)}\frac{\tau_{5/2}^2}{(1+\tau_{5/2}^2)} \Big \} \ ,
\een
where $\theta_i=\hbar\omega_{c_i}/k_BT$, $\tau_{i,1}=\tau_{i,0}/(1+\theta_i)$, $\tau_{i,2}=\tau_{i,0}/(1+2\theta_{i,2})$, and $\tau_{i,h}=\tau_{i,0}/(1+5\theta_i/2)$.

\subsection{Expressions for transfer rates in the GME-MED}
Within the second order approximation with respect to $J_{sn,qm}^c$'s, 
\ben
&&W_{1\rightarrow 2}(t)=\frac{2}{\hbar^2}{\rm Re}\sum_{n,n',m,m'=1}^9 \sum_{s,s',q,q'=\alpha}^{\beta} J_{sn,qm}^{c}J_{sn',q'm'}^{c} \nonumber \\
&&\hspace{.4 in}\int_0^t d\tau \langle q_{2,m}|{\mathcal I}_2 (\tau) |q_{2,m'}'\rangle \langle s_{1,n'}'| {\mathcal E}_{1}^s(\tau)|s_{1,n}\rangle \ . \label{eq:k_nmt_gme}
\een
$W_{2\rightarrow 1}(t)$ has the same expression as above except that $1$ and $2$ are interchanged with each other.
In the limit of $t\rightarrow \infty$, the $W_{1\rightarrow 2}(t)$ ($W_{2\rightarrow 1}(t)$) approaches the multichromophoric F\"{o}rster resonance energy transfer (MC-FRET) rate\cite{jang-prl92} from $1$ to $2$ ($2$ to $1$).

For the line shape expressions entering Eq. (\ref{eq:k_nmt_gme}), an approximation that neglects off-diagonal components in the exciton basis is used as detailed below.   
For this, denote the  $p$th eigenstate of $H_{e,k}^0$ (with $k=1$ or $2$) with energy $\epsilon_{p_k}$ as $|\varphi_{p_k}\rangle$, and 
define the following transformation matrix:
\be
U_{k,s_np}=\langle s_{k,n}|\varphi_{p_k}\rangle\ .
\ee
Let us also introduce
\ben
&& \lambda_{p_k}=(\sum_{s=\alpha}^\beta\sum_{n=1}^9 |U_{k,s_{n}p}|^4 ) \lambda \ , \\
&&G_{p_k,i}(t)=(\sum_{s=\alpha}^\beta \sum_{n=1}^9 |U_{k,s_np}|^4 ) G_i(t) \ , \\
&&G_{p_k,r}(t)=(\sum_{s=\alpha}^\beta\sum_{n=1}^9 |U_{k,s_np}|^4 ) G_r(t) \ , 
\een  
where $\lambda$, $G_i(t)$, and $G_r(t)$ are defined by Eqs. (\ref{eq:lambda})-(\ref{eq:gr}).
Then, within the approximation that considers only the diagonal elements in the exciton states, Eq. (\ref{eq:k_nmt_gme}) can be expressed as
\ben 
&&W_{1\rightarrow 2} (t)=\frac{2}{\hbar^2} {\rm Re} \sum_{p_1}\sum_{p_2'}\frac{e^{-\tilde \epsilon_{p_1}/k_BT} }{(\sum_{p_1''} e^{- \tilde \epsilon_{p_1''}/k_BT})} |\tilde J_{p_1p_2'}^c|^2\nonumber \\
&&\hspace{0.6in}\times \int_0^t d\tau  e^{-G_{p_2',r}(\tau)-iG_{p_2',i}(\tau)-i\tilde \epsilon_{p_2'}\tau/\hbar} \nonumber \\
&&\hspace{0.8in}\times   e^{-G_{p_1,r}(\tau)-iG_{p_1,i}(\tau)+i\tilde \epsilon_{p_1} \tau/\hbar} \ , \label{eq:knmt-exciton}
\een
where
\ben
&&\tilde \epsilon_{p_k}=\epsilon_{p_k}-\lambda_{p_k} \ ,\\
&&\tilde J_{p_1p_2'}^c=\sum_{s,q=\alpha}^\beta \sum_{n,m=1}^9 U_{2,q_mp'} J_{sn,qm}^c U^*_{1,s_np} \ .
\een
The expression for $W_{2\rightarrow 1}(t)$ is the same as Eq. (\ref{eq:knmt-exciton}) except that $1$ and $2$ are interchanged with each other.
\subsection{S3. Comparison of real time population dynamics}
Figure S1 compares the population dynamics based on the GME-MED method with the exponential dynamics based on the MC-FRET rate (left column) and also the population dynamics averaged over the ensemble of disorder.  Three inter-LH2 distances were considered.  The deviation of the GME-MED dynamics from the MC-FRET data on the left column indicates that the population dynamics is non-exponential and makes the effective GME-MED rate slower than the MC-FRET rate.    The almost perfect agreement of the ensemble averaged population dynamics on the right hand side is consistent with the fact that the corresponding average GME-MED rate becomes virtually the same as the average MC-FRET rate.  
\ \vspace{.5in}\\
\begin{figure}
\includegraphics[width=5in]{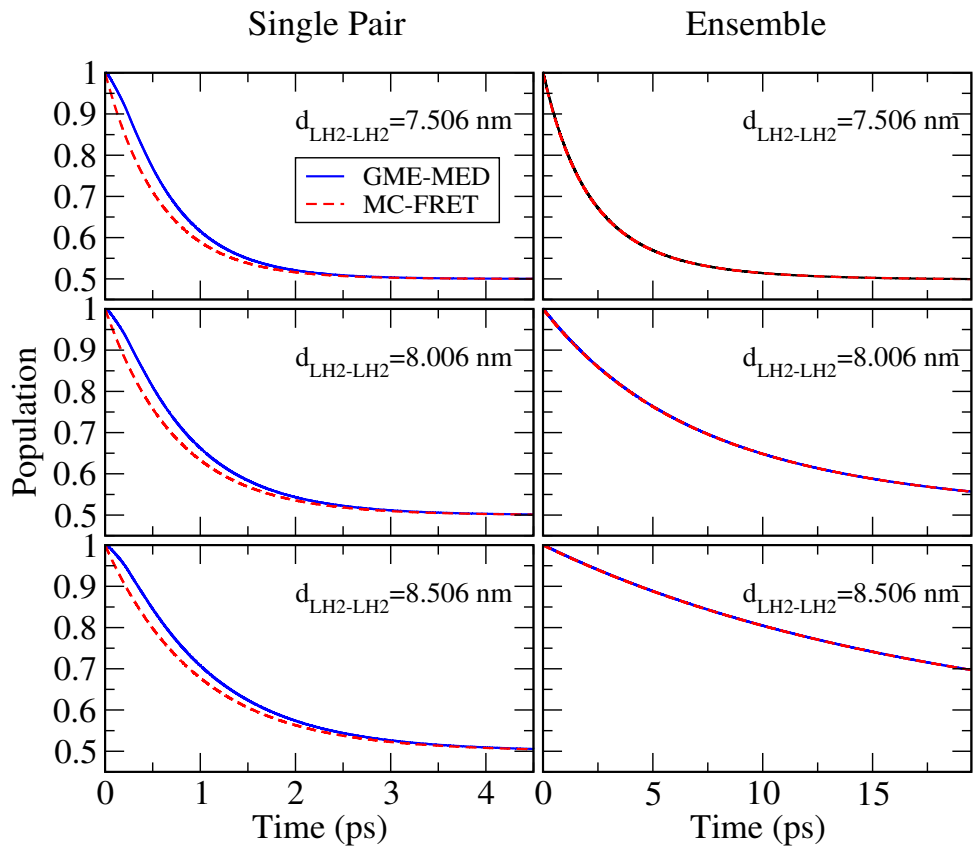} 
\caption{Time dependent population of the exciton in the initial B850 unit calculated from the GME-MED method (blue solid line) and the exponential decay based on the MC-FRET rate (red dashed line).  The left column is for the case where there is no disorder.  The right column shows the average population dynamics for the entire ensemble of the disorder as specified in the text.} 
\end{figure}
\ \vspace{.5in}\\

\providecommand{\latin}[1]{#1}
\providecommand*\mcitethebibliography{\thebibliography}
\csname @ifundefined\endcsname{endmcitethebibliography}
  {\let\endmcitethebibliography\endthebibliography}{}

\end{widetext}

\end{document}